\documentclass[a4paper,10pt]{article}

\usepackage[utf8]{inputenc}
\usepackage[spanish,english]{babel}
\usepackage{graphicx}
\usepackage{amssymb}
\usepackage{amsmath}
\usepackage{hyperref}
\usepackage{color}
\usepackage{latexsym,dsfont,mathtools}
\usepackage{theorem,titletoc,titlesec}
\usepackage{enumitem}
\usepackage{tikz}
\usepackage{tabularx}
\usetikzlibrary{matrix}
\usepackage{adjustbox}
\newtheorem{theorem}{Theorem}[section]
\newtheorem{corollary}{Corollary}[section]

\newtheorem{example}{Example}[section]
\newtheorem{remark}{Remark}[section]

\usepackage{multirow}
\usepackage{tabularx}
\usepackage{xcolor}

\newcommand{\Z}{\mathbb{Z}}

\newcommand{\C}{{\cal C}}

\newcommand{\wt}{{\rm wt}}

\newcommand{\rank}{\text{rank}}
\newcommand{\kernel}{\text{ker}}

\newcommand{\cH}{\cal{H}}

\begin{document}

\title{$\Z_p\Z_{p^2}\dots\Z_{p^s}$-Additive Generalized Hadamard Codes\thanks{This work has been partially supported by the Spanish Ministerio de Ciencia e Innovación under Grant PID2019-104664GB-I00 (AEI / 10.13039/501100011033) and by the Catalan AGAUR scholarship 2020 FI SDUR 00475.}%\thanks{The material in this paper was presented in part at the ...}
}

\author{Dipak Kumar Bhunia, Cristina Fern\'andez-C\'ordoba, Merc\`e Villanueva\thanks{The authors are with the Department of Information and Communications
Engineering, Universitat Aut\`{o}noma de Barcelona, 08193 Cerdanyola del Vall\`{e}s, Spain.}}

%  \IEEEauthorblockA{Universitat Aut\`{o}noma de Barcelona \\ 
%                    08193 Cerdanyola del Vall\`{e}s, Spain \\
%                    Email: \{cristina.fernandez, carlos.vela, merce.villanueva\}@uab.cat}


\maketitle

%%%%%%
%% Abstract: 
%% If your paper is eligible for the student paper award, please add
%% the comment "THIS PAPER IS ELIGIBLE FOR THE STUDENT PAPER
%% AWARD." as a first line in the abstract. 
%% For the final version of the accepted paper, please do not forget
%% to remove this comment!
%%
\begin{abstract}
The $\Z_p\Z_{p^2}\dots\Z_{p^s}$-additive codes are subgroups of $\Z_p^{\alpha_1} \times \Z_{p^2}^{\alpha_2} \times \cdots \times \Z_{p^s}^{\alpha_s}$, and can be seen as linear codes over $\Z_p$ when $\alpha_i=0$ for all  $i \in \{2,\dots, s\}$,   $\Z_{p^s}$-additive codes when $\alpha_i=0$ for all $i \in \{1,\dots, s-1\}$ , or a $\Z_p\Z_{p^2}$-additive code when $s=2$, %$\alpha_i=0$ for all $i \in \{3,4,\dots, s\}$, 
or $\Z_2\Z_4$-additive codes when $p=2$ and $s=2$. A $\Z_p\Z_{p^2}\dots\Z_{p^s}$-linear generalized Hadamard (GH) code is a GH code over $\Z_p$ which is the Gray map image of a $\Z_p\Z_{p^2}\dots\Z_{p^s}$-additive code.  
In this paper, we generalize some known results for  $\Z_p\Z_{p^2}\dots\Z_{p^s}$-linear GH codes with $p$ prime and $s\geq 2$. First, we give a recursive construction of  $\Z_p\Z_{p^2}\dots\Z_{p^s}$-additive GH codes of type $(\alpha_1,\dots,\alpha_s;t_1,\dots,t_s)$ with $t_1\geq 1, t_2,\dots,t_{s-1}\geq 0$, and $t_s\geq1$. Then, we show  for which types the corresponding $\Z_p\Z_{p^2}\dots\Z_{p^s}$-linear GH codes  are  nonlinear over $\Z_p$. We also compute the kernel and its dimension whenever they are nonlinear. 

%PAPER: Moreover,  we show that, unlike $\Z_4$-linear GH codes, the $\Z_{p^2}$-linear GH codes are not included in the family of $\Z_p\Z_{p^2}$-linear GH codes with $\alpha_1\not =0$ when $p\geq 3$ prime. Actually, none of these $\Z_p\Z_{p^2}$-linear GH codes is equivalent to a $\Z_{p^s}$-linear GH code with $s\geq 2$. 
\end{abstract}

%\end{titlepage}

%\newpage
%\tableofcontents
%\newpage

\section{Introduction}
Let $\Z_{p^s}$ be the ring of integers modulo $p^s$ with $p$ prime and $s\geq1$. The set of
$n$-tuples over $\Z_{p^s}$ is denoted by $\Z_{p^s}^n$. In this paper,
the elements of $\Z^n_{p^s}$ will also be called vectors. 
%The order of a vector $\mathbf u$ over $\Z_{p^s}$, denoted by $o(\mathbf{u})$, is the smallest positive integer $m$ such that $m \mathbf{u} =\zero$.
A code over $\Z_p$ of length $n$ is a nonempty subset of $\Z_p^n$,
and it is linear if it is a subspace of $\Z_{p}^n$. Similarly, a nonempty
subset of $\Z_{p^s}^n$ is a $\Z_{p^s}$-additive if it is a subgroup of $\Z_{p^s}^n$. A $\Z_p\Z_{p^2}\dots\Z_{p^s}$-additive code is a subgroup of $\Z_p^{\alpha_1} \times \Z_{p^2}^{\alpha_2} \times \cdots \times \Z_{p^s}^{\alpha_s}$ for some non-negative integers $\alpha_1,\ldots,\alpha_s$. 
Note that a $\Z_p\Z_{p^2}\dots\Z_{p^r}$-additive code is a linear code over $\Z_p$ when $\alpha_i=0$ for all  $i \in \{2,3,\dots, s\}$,  a $\Z_{p^r}$-additive code, where $r\in \{2,\dots,s\}$, when $\alpha_i=0$ for all $i \in \{1,2,\dots, s\} \setminus \{r\}$, or a $\Z_p\Z_{p^2}$-additive code when $\alpha_i=0$ for all $i \in \{3,4,\dots, s\}$. The order of a vector $u\in \Z_{p^s}$, denoted by $o(u)$, is the smallest positive integer $m$ such that $m u =(0,\dots,0)$. Also, the order of $\mathbf u\in \Z_p^{\alpha_1}\times\Z_{p^2}^{\alpha_2} \times \cdots \times \Z_{p^s}^{\alpha_s} $, denoted by $o(\mathbf u)$, is the smallest positive integer $m$ such that $m \mathbf u =(0,\dots,0\mid \cdots \mid  0,\dots,0)$.

The Hamming weight of a vector $u\in\Z_{p}^n$, denoted by $\wt_H(u)$, is
the number of nonzero coordinates of $u$. The Hamming distance of two
vectors $u,v\in\Z_{p}^n$, denoted by $d_H(u,v)$, is the number of
coordinates in which they differ.  Note that $d_H(u,v)=\wt_H(u-v)$. The minimum distance of a code $C$ over $\Z_p$ is $d(C)=\min \{ d_H(u,v) : u,v \in C, u \not = v  \}$.

In \cite{Sole}, a Gray map  from $\Z_4$ to $\Z_2^2$ is defined as
$\phi(0)=(0,0)$, $\phi(1)=(0,1)$, $\phi(2)=(1,1)$ and $\phi(3)=(1,0)$. There exist different generalizations of this Gray map, which go from $\Z_{2^s}$ to
$\Z_2^{2^{s-1}}$ \cite{Carlet,Codes2k,dougherty,Nechaev,Krotov:2007}.
The one given in \cite{Nechaev} can be defined in terms of the elements of a Hadamard code \cite{Krotov:2007}, and Carlet's Gray map \cite{Carlet} is a particular case of the one given in \cite{Krotov:2007} 
satisfying $\sum_{i=0}^{s-1} \lambda_i \phi(2^i) =\phi(\sum_{i=0}^{s-1} \lambda_i 2^i)$ \cite{KernelZ2s}. 
In this paper, we focus on a generalization of Carlet's Gray map, from $\Z_{p^r}$ to $\Z_p^{p^{r-1}}$, which is also a particular case of the one given in \cite{ShiKrotov2019}. Specifically, \begin{gather*}\label{eq:GrayMapCarlet}
\phi_{r}(u)=(u_{r-1},\dots,u_{r-1})+(u_0,\dots,u_{r-2})Y_{r-1},
\end{gather*}
where $u\in\Z_{p^r}$; $[u_0,u_1,\dots,u_{r-1}]_p$ is the $p$-ary expansion of $u$, that is, $u=\sum_{i=0}^{r-1}u_ip^i$ with $u_i\in \Z_p$; and $Y_{r-1}$ is a matrix of size $(r-1)\times p^{r-1}$ whose columns are all the vectors in $\Z_p^{r-1}$. Without loss of generality, we assume that the columns of $Y_{r-1}$ are ordered in ascending order, by considering the elements of $\mathbb{Z}_p^{r-1}$ as the $p$-ary expansions of the elements of $\mathbb{Z}_{p^{r-1}}$. Note that $\phi_1$ is the identity map.
We define $\Phi_{r}:\Z_{p^r}^n \rightarrow \Z_p^{np^{r-1}}$ as the component-wise extended map of $\phi_{r}$.
We can also define a Gray map $\Phi$ from $\Z_p^{\alpha_1} \times \Z_{p^2}^{\alpha_2} \times \cdots \times \Z_{p^s}^{\alpha_s}$ to $\Z_p^n$, where $n=\alpha_1+p\alpha_2+\cdots +p^{s-1}\alpha_s$, as follows:
$$
\Phi(y_1\mid y_2\mid \dots \mid y_s)=(y_1, \Phi_2(y_2), \dots,\Phi_s(y_s)),
$$
for any   $y_i \in \Z_{p^i}^{\alpha_i}$, where $1\leq i\leq s$.

%Let $\Phi:\Z_p^{\alpha_1} \times \Z_{p^2}^{\alpha_2} \times \cdots \times \Z_{p^s}^{\alpha_s} \rightarrow\Z_p^n$, where $n=\alpha_1+p\alpha_2+\cdots +p^{s-1}\alpha_s$, be an extension of the Gray map $\phi$ given by 
%$$
%\Phi(\mathbf{x}\mid \mathbf{y}_1\mid \dots \mid \mathbf{y}_{s-1})=(\mathbf{x}, \Phi_1(y_1), \dots,\Phi_{s-1}(y_{s-1})),
%$$
%for any $\mathbf{x} \in \Z_p^{\alpha_1}$ and  $\mathbf{y}_i \in \Z_{p^{i+1}}^{\alpha_{i+1}}$, where $1\leq i\leq s-1$.

%$$
%\Phi(\mathbf{x}\mid \mathbf{y}_1\mid \dots \mid \mathbf{y}_{s-1})=(\mathbf{x}, \phi_1(y_{11}), \dots, \phi_1(y_{1\alpha_2}),\dots,  \phi_{s-1}(y_{s-1,1}),\dots, \phi_{s-1}(y_{s-1,\alpha_s})),
%$$
%for any $\mathbf{x} \in \Z_p^{\alpha_1}$ and  $\mathbf{y}_i=(y_{i1},\dots,y_{i\alpha_{i+1}}) \in \Z_{p^{i+1}}^{\alpha_{i+1}}$, where $1\leq i\leq s-1$.

Let $\C$ be a $\Z_p\Z_{p^2}\dots\Z_{p^s}$-additive code. We say that its Gray map image $C=\Phi(\C)$ is a $\Z_p\Z_{p^2}\dots\Z_{p^s}$-linear code of length $n=\alpha_1+p\alpha_2+\cdots +p^{s-1}\alpha_s$.
Since $\C$ is a subgroup of
$\Z_p^{\alpha_1} \times \Z_{p^2}^{\alpha_2} \times \cdots \times \Z_{p^s}^{\alpha_s}$, it is isomorphic to an abelian structure
$\Z_{p^s}^{t_1}\times\Z_{p^{s-1}}^{t_2}\times
\dots \times\Z_p^{t_s}$, and we say that $\C$, or equivalently
$C=\Phi(\C)$, is of type $(\alpha_1,\dots, \alpha_s;t_1,\dots,t_{s})$.
Note that $|\C|=p^{st_1}p^{(s-1)t_2}\cdots p^{t_s}$.
Unlike linear codes over finite fields,
linear codes over rings do not have a basis, but there
exists a generator matrix for these codes having minimum number of rows, that is, $t_1+\cdots+t_s$ rows.  

Two structural properties of codes over $\Z_p$ are the rank and
dimension of the kernel. The rank of a code $C$ over $\Z_p$ is simply the
dimension of the linear span, $\langle C \rangle$,  of $C$.
The kernel of a code $C$ over $\Z_p$ is defined as
$\mathrm{K}(C)=\{\textbf{x}\in \Z_p^n : \textbf{x}+C=C \}$ \cite{BGH83,pKernel}. If the all-zero vector belongs to $C$,
then $\mathrm{K}(C)$ is a linear subcode of $C$.
Note also that if $C$ is linear, then $K(C)=C=\langle C \rangle$.
We denote the rank of $C$ as $\rank(C)$ and the dimension of the kernel as $\kernel(C)$.
These parameters can be used to distinguish between non-equivalent codes, since equivalent ones have the same rank and dimension of the kernel.

%A binary code of length $n$, $2n$ codewords and minimum distance $n/2$ is called a Hadamard code. Hadamard codes can be constructed from Hadamard matrices \cite{Key,WMcwill}. Note that linear Hadamard codes are in fact first order Reed-Muller codes, or equivalently, the dual of extended Hamming codes \cite[Ch.13 \S 3]{WMcwill}.

%One can follow paper \cite{jungnickel1979} for the generalized Hadamard matrices and \cite{dougherty2015ranks} for the corresponding generalized Hadamard codes. 

A generalized Hadamard $(GH)$ matrix $H(p,\lambda) = (h_{i j})$ of order $N = p\lambda$ over $\Z_p$ is a $p\lambda \times p\lambda$ matrix with entries from $\Z_p$ with the property that for every $i, j$, $1 \leq i < j \leq p\lambda,$ each of the multisets $\{h_{is}- h_{js} : 1 \leq s \leq p\lambda\}$ contains every element of $\Z_p$ exactly $\lambda$ times \cite{jungnickel1979}. 
%It is known that since $(\Z_p , +)$ is an abelian group, then ${H(p,\lambda)}^T$ is also a $GH$ matrix, where ${H(p,\lambda)}^T$ denotes the transpose of $H(p,\lambda)$ \cite{jungnickel1979,}. 
An ordinary Hadamard matrix of order $4\mu$ corresponds to  $GH$ matrix $H(2,\lambda)$ over $\Z_2$, where $\lambda = 2\mu$ \cite{Key}. 
Two $GH$ matrices $H_1$ and $H_2$ of order $N$ are said to be equivalent if one can be obtained from the other by a permutation of the rows and columns and adding the same element of $\Z_p$ to all the coordinates in a row or in a column. 

We can always change the first row and column of a $GH$ matrix into zeros and we obtain an equivalent $GH$ matrix which is called normalized. From a normalized GH matrix $H$, we denote by $F_H$ the code consisting of the rows of $H$, and $C_H= \bigcup_{\alpha \in\Z_p} (F_H + \alpha \textbf{1})$,
where $F_H + \alpha \textbf{1} = \{h + \alpha \textbf{1} : h \in F_H\}$ and $\textbf{1}$ denotes the
all-one vector. The code $C_H$ over $\Z_p$ is called generalized
Hadamard $(GH)$ code \cite{dougherty2015ranks}. Note that $C_H$ is generally a nonlinear code over $\Z_p$. Moreover, if it is of length $N$, it has $pN$ codewords and minimum distance $N(p-1)/p$.

The $\Z_p\Z_{p^2}\dots\Z_{p^s}$-additive codes such that after the Gray map $\Phi$ give
GH codes are called $\Z_p\Z_{p^2}\dots\Z_{p^s}$-additive GH codes and the
corresponding images are called $\Z_p\Z_{p^2}\dots\Z_{p^s}$-linear GH codes. 
It is known that $\Z_2\Z_4$-linear GH codes with $\alpha_1=0$ and $\alpha_1\not =0$ can be classified by using either the rank or the dimension of the kernel \cite{Kro:2001:Z4_Had_Perf,PRV06}. In \cite{KV2015}, it is shown that each $\Z_2\Z_4$-linear GH code with $\alpha_1 =0$ is  equivalent to a  $\Z_2\Z_4$-linear GH code with $\alpha_1 \not =0$, so indeed there are only $\lfloor
t/2\rfloor$ non-equivalent $\Z_2\Z_4$-linear GH codes of length $2^t$.
Later, in \cite{KernelZ2s,HadamardZps}, an iterative construction for $\Z_{p^s}$-linear GH codes is described, the linearity is established, and a partial classification by using the dimension of the kernel is obtained,
giving the exact amount of non-equivalent such codes for some parameters. An iterative construction for $\Z_p\Z_{p^2}$-linear GH codes with $\alpha_1\neq 0$ is described in \cite{ZpZp2Construction,WCC2022}.

 This paper is focused on $\Z_p\Z_{p^2}\dots\Z_{p^s}$-linear GH codes with $p$ prime and $s\geq2$, generalizing some results related to the construction, linearity, and kernel of such codes  given in \cite{PRV06,RSV08, ZpZp2Construction,WCC2022}. 
This paper is organized as follows.
In Section \ref{sec:Construction}, we describe a recursive construction of  $\Z_p\Z_{p^2}\dots \Z_{p^s}$-linear GH codes of type $(\alpha_1,\dots, \alpha_s;t_1,\dots,t_s)$.
In Sections \ref{Sec:Linearity}, we establish for which types these codes are linear, and we give the kernel and its dimension whenever they are nonlinear. 
%We also show that the dimension of the kernel is enough to classify completely the $\Z_p\Z_{p^2}$-linear GH codes with $\alpha_1\not =0$ of a given length, providing the number of non-equivalent such codes, like it was proved for $\Z_2\Z_4$-linear GH codes in \cite{PRV06}.

\section{Construction of $\Z_p\Z_{p^2}\dots \Z_{p^s}$-additive GH codes}\label{sec:Construction}

Let  $\mathbf{0}, \mathbf{1},\mathbf{2},\ldots, \mathbf{p^s-1}$ be the vectors having the elements $0, 1, 2, \ldots, p^s-1$  repeated in each coordinate, respectively, where $p$ is a prime number. Let
\begin{equation}\label{eq:ZpsRecGenMatrix0}
A_p^{1,0,\dots,0,1}=
\left(\begin{array}{ccc|ccc|c|ccc}
1  &\cdots &1  &p  &\cdots &p &\cdots &p^{s-1}  &\cdots &p^{s-1} \\
0   &\cdots &p-1  &1 &\cdots &p-1 &\cdots &1  &\cdots &p-1 \\
\end{array}\right).
\end{equation}
Let $t_1\geq 1$, $t_2,\dots, t_{s-1}\geq 0$, and $t_s\geq 1$ be integers. Any matrix $A_p^{t_1, \dots, t_s}$ with $(t_1,\dots,t_s) \neq (1,0,\dots,0,1)$ is constructed recursively starting from $A_p^{1,0,\dots,0,1}$ in the following way. First, if $A$ is a generator matrix of a $\Z_p\Z_{p^2}\dots \Z_{p^s}$-additive code, that is, a subgroup of $\Z_p^{\alpha_1} \times \Z_{p^2}^{\alpha_2} \times \cdots \times \Z_{p^s}^{\alpha_s}$, then we denote by $A_i$ the submatrix of $A$ consisting of the columns of A which are over $\Z_{p^i}$ for $i\in \{1,\dots,s\}$. We have that $A=(A_1\mid\cdots\mid A_s)$, where the number of columns of $A_i$ is $\alpha_i$.  To achieve $A_p^{t_1, \dots,t_s}$, we apply construction (\ref{eq:ZpsRecGenMatrix1}) $t_1+t_2+\cdots+t_s -2$ times to add the rows to the matrix in the following order. First, we start with the matrix $A_p^{1,0,\dots,0,1}$ and we add $t_1-1$ rows of order $p^s$, up to obtain $A_p^{t_1,0,\dots,0,1}$; then $t_2$ rows of order $p^{s-1}$ up to generate $A_p^{t_1,t_2,0,\dots,0,1}$; and  so on, until we add $t_{s}-1$ rows of order $p$ to achieve  $A_p^{t_1, \dots,t_s}$. %However, for $s=2$, one can also follow the construction given in \cite{ZpZp2Construction}.
Now, the construction is as follows.
Let $A=A_p^{t_1,\dots,t_s}=(A_1\mid\dots\mid A_s)$, and $i\in\{1,\dots,s\}$. If $i<s$, then we consider the matrices 
$$P_j=
\left(\begin{array}{cccccc}
M_j &\cdots &M_j &A_{j+1} &\cdots &A_{j+1}\\
\mathbf{1} &\cdots & \mathbf{p-1} &\mathbf{0}  &\cdots & \mathbf{p^{j+1}-1}\\
\end{array}\right),$$
%\end{equation*}
where $M_j=\{\mathbf{z}^T: \mathbf{z}\in\lbrace p^j \rbrace\times \lbrace p\cdot 0,p\cdot 1,\dots, p\cdot(p^j-1)\rbrace^{t_1+\cdots+t_s-1}\}$ %\times \stackrel{(t_1+\cdots+t_i-1)}{\dots}\times \lbrace p\cdot 0,p\cdot 1,\cdots, p\cdot(p^j-1) \rbrace\}$ 
for $j\in \{1,\dots, s-i\}$. If $i>1$, we consider 
$$
Q_k=
\left(\begin{array}{ccc}
A_{s-i+k+1} &\cdots &A_{s-i+k+1}\\
p^k\cdot \mathbf{0}  &\cdots & p^k\cdot \mathbf{p^{s-i+1}-1}\\
\end{array}\right)
$$
for $k\in \{1,\dots, i-1\}$.
Now, we construct the matrix $A_p^{t'_1,\cdots, t'_s}$, with $t'_i=t_i+1$  and $t'_\ell=t_\ell$ for $\ell\not=i$ as
\begin{equation}\label{eq:ZpsRecGenMatrix1}
\left(\begin{array}{@{}c|c|c|c|c|c|c}
\begin{matrix}
      A_1 &\cdots &A_1\\
      \mathbf{0}  &\cdots & \mathbf{p-1}\\
   \end{matrix}
   &P_1 &\cdots &P_{s-i} &Q_1 &\cdots &Q_{i-1}\\
   \end{array}\right).
\end{equation}

% Along this paper, we consider that the matrices $A_p^{t_1,t_2, \dots, t_s}$ are constructed recursively starting from $A_p^{1,0,\dots,0,1}$ in the following way. First, we add $t_1-1$ rows of order $p^s$, up to obtain $A_p^{t_1,0,\dots,0,1}$; then $t_2$ rows of order $p^{s-1}$ up to generate $A_p^{t_1,t_2,0,\dots,0,1}$; and  so on, until we add $t_s-1$ rows of order $p$ to achieve  $A_p^{t_1,t_2, \dots, t_s}$.

\begin{example}
Let $p=2$, $s=3$, and  $\mathbf{0}, \mathbf{1},\mathbf{2},\ldots, \mathbf{7}$ be the vectors having the elements $0, 1, 2, \ldots, 7$  repeated in each coordinate, respectively. Let
\begin{equation}\label{eq:recGenMatrix00}
A_2^{1,0,1}=
\left(\begin{array}{cc|c|c}
1 & 1  & 2  &4 \\
0  & 1 &1  &1  \\
\end{array}\right).%=\left(\begin{array}{c|c|c} A_1 & A_2 & A_3 \\ \end{array}\right).
\end{equation}
Let $t_1\geq 1$, $t_2\geq 0$, and $t_3\geq 1$ be integers. Suppose, we want to construct $A_2^{t_1,t_2,t_3}$, where $(t_1,t_2,t_3)\neq (1,0,1)$. We start with the matrix $A_2^{1,0,1}$. If we have matrix  $A_2^{\ell-1,0,1}=(A_1 \mid A_2 \mid A_3)$, $\ell \geq 2$, we may construct the matrix
\begin{equation}\label{eq:recGenMatrix01}
\footnotesize
A_2^{\ell,0,1}=
\left(\begin{array}{cc|ccccc|ccccc}
A_1 & A_1 &M_1 &A_2 &A_2 &A_2 &A_2 &M_2 &A_3 &A_3 &\cdots &A_3 \\
\mathbf{0}  & \mathbf{1} & \mathbf{1}  &\mathbf{0} &\mathbf{1}  &\mathbf{2} &\mathbf{3} &\mathbf{1} &\mathbf{0} &\mathbf{1} &\cdots &\mathbf{7} \\
\end{array}\right),
\end{equation}
where $M_1=\{\mathbf{z}^T:\mathbf{z}\in\lbrace2\rbrace\times \lbrace0,2\rbrace^{\ell-1}\}$ and $M_2=\{\mathbf{z}^T: \mathbf{z}\in\lbrace4\rbrace\times \lbrace0,2,4,6\rbrace^{\ell-1}\}$. For example, from $A_2^{1,0,1}$, we have 
\begin{equation} \label{eq:A2201}
    A_2^{2,0,1}=
\left(\begin{array}{cc|cc|cc}
11 &11  &22 &2222 &4444 &44444444 \\
01 &01  &02 &1111 &0246 &11111111 \\
00 &11  &11 &0123 &1111 &01234567\\
\end{array}\right).
\end{equation}
We perform construction (\ref{eq:recGenMatrix01}) until $\ell=t_1$. If we have matrix $A_2^{t_1,\ell-1,1}=(A_1\mid A_2 \mid A_3)$, $t_1 \geq 1$, $\ell\geq 1$, we may construct the matrix 
\begin{equation}\label{eq:recGenMatrix02}
\footnotesize
A_2^{t_1,\ell,1}=
\left(\begin{array}{cc|ccccc|cccc}
A_1 & A_1 &M_1 &A_2 &A_2 &A_2 &A_2 &A_3 &A_3 &A_3 &A_3 \\
\mathbf{0}  & \mathbf{1} & \mathbf{1}  &\mathbf{0} &\mathbf{1}  &\mathbf{2} &\mathbf{3}  &\mathbf{0} &\mathbf{2} &\mathbf{4} &\mathbf{6} \\
\end{array}\right),
\end{equation}
where $M_1=\{\mathbf{z}^T:\mathbf{z}\in\lbrace2\rbrace\times \lbrace0,2\rbrace^{t_1+\ell-1}\}$. For example, if $A_2^{2,0,1}=(A_1\mid A_2\mid A_3)$ is the matrix given in (\ref{eq:A2201}), we can construct the matrix 
\begin{equation}
A_2^{2,1,1}=\left(\begin{array}{cc|@{}cccccc|cccc}
      A_1  &A_1 &&\begin{matrix}
            22 2 2\\
            0 0 2 2\\
            0 2 0 2\\
      \end{matrix} &A_2  &A_2  &A_2  &A_2 &A_3 &A_3 &A_3 &A_3\\
      \mathbf{0} &\mathbf{1} &&\mathbf{1} &\mathbf{0} &\mathbf{1} &\mathbf{2} &\mathbf{3} &\mathbf{0} &\mathbf{2} &\mathbf{4} &\mathbf{6} \\
   \end{array}\right).
\end{equation}
We repeat construction (\ref{eq:recGenMatrix02}) until $\ell=t_2$. If we have matrix $A_2^{t_1,t_2,\ell-1}=(A_1\mid A_2 \mid A_3)$, $t_1\geq 1$, $t_2\geq 0$, and $\ell\geq 2$, we may construct the matrix
\begin{equation}\label{eq:recGenMatrix03}
A_2^{t_1,t_2,\ell}=
\left(\begin{array}{cc|cc|cc}
A_1 & A_1 & A_2 & A_2 &A_3 &A_3 \\
\mathbf{0}  & \mathbf{1} & \mathbf{0} &\mathbf{2}  &\mathbf{0}  &\mathbf{4}  \\
\end{array}\right).
\end{equation}
We repeat construction (\ref{eq:recGenMatrix03}) until $\ell=t_3$. That way, we obtain $A_2^{t_1,t_2,t_3}$.
\end{example}

\medskip
The $\Z_p\Z_{p^2}\dots\Z_{p^s}$-additive code generated by $A_p^{t_1,\dots,t_s}$ is denoted by ${\cH}_p^{t_1,\dots,t_s}$, and the corresponding $\Z_p\Z_{p^2}\dots\Z_{p^s}$-linear code $\Phi( {\cH}_p^{t_1,\dots,t_s})$ by $H_p^{t_1,\dots,t_s}$.

\begin{theorem}
The $\Z_p\Z_{p^2}\dots\Z_{p^s}$-additive code ${\cH}_p^{t_1,\dots,t_s}$ generated by the matrix
$A_p^{t_1,\dots,t_s}$ is a $\Z_p\Z_{p^2}\dots\Z_{p^s}$-additive GH code of type $(\alpha_1,\dots, \alpha_s;t_1,\dots,t_{s})$ with $t_1\geq 1, t_2,\dots,t_{s-1}\geq 0$  and $t_s\geq1$.
\end{theorem}

\begin{remark}
For $s=2$, we know from \cite{ZpZp2Construction} that we do not need to follow exactly the above order to obtain $A_p^{t_1,t_2}$ from $A_p^{1,1}$. However, for $s\geq 3$, it is mandatory to follow exactly the above order; otherwise, it may not give a GH code.
\end{remark}

\section{Linearity and kernel of $\Z_p\Z_{p^2}\dots \Z_{p^s}$-linear GH codes}\label{Sec:Linearity} 
In this section, we generalize the results about linearity and kernel of $\Z_p\Z_{p^2}$-linear GH codes given in \cite{PRV06,WCC2022}. The first result shows that the codes $H_2^{1,0,\dots,0,t_s}$, with $t_s\geq 1$, are the only $\Z_2\Z_{2^2}\dots \Z_{2^s}$-linear Hadamard codes which are linear. 
Then, the next result shows that, for $p\geq 3$ prime, there are no $\Z_p\Z_{p^2}\dots\Z_{p^s}$-linear GH codes of type $(\alpha_1,\dots, \alpha_s;t_1,\dots,t_{s})$ with $t_1\geq 1$, $t_2,\dots,t_{s-1}\geq 0$,  and $t_s\geq1$, which are linear. When these codes are nonlinear, we also give the results about the kernel and a basis of the kernel, which gives us its dimension. Specifically, we see that the dimension of the kernel of a nonlinear $\Z_p\Z_{p^2}\dots\Z_{p^s}$-linear GH code of type $(\alpha_1,\dots, \alpha_s;t_1,\dots,t_{s})$ with $p$ prime is $t_1+\dots+t_s$.

\begin{theorem}
The $\Z_2\Z_{2^2}\dots\Z_{2^s}$-linear Hadamard codes $H_2^{1,0,\dots,0,t_s}$, with $t_s\geq 1$, are the only $\Z_2\Z_{2^2}\dots\Z_{2^s}$-linear Hadamard codes which are linear.
\end{theorem}
\begin{theorem}
   Let $t_1\geq 1$, $t_2,\dots,t_{s-1}\geq 0$, and $t_s\geq1$ be integers.
   Let ${\cH}={\cH}_p^{t_1,t_2,\dots,t_s}$ be the $\Z_p\Z_{p^2}\dots\Z_{p^s}$-additive GH code of type $(\alpha_1,\dots, \alpha_s;t_1,\dots,t_{s})$ with $p\geq 3$ prime. Then, $\Phi(\mathcal{H})$ is nonlinear.
\end{theorem}

%\section{Kernel of $\Z_p\Z_{p^2}\dots \Z_{p^s}$-linear GH codes}\label{Sec:Kernel}
\begin{theorem}
   Let $t_1\geq 1$, $t_2,\dots,t_{s-1}\geq 0$, and $t_s\geq1$ be integers.
   Let  ${\cH}={\cH}_p^{t_1,t_2,\dots,t_s}$ be the $\Z_p\Z_{p^2}\dots\Z_{p^s}$-additive GH code of type $(\alpha_1,\dots, \alpha_s;t_1,\dots,t_{s})$, with $p$ prime, such that $\Phi({\cH})$ is nonlinear. Let $\mathcal{H}_p$ be the subcode of $\mathcal{H}$ which contains all the codewords of order atmost $p$. Then, $K(\Phi(\mathcal{H}))=\Phi(\mathcal{H}_p)$.
\end{theorem}

\begin{corollary}\label{coro:kernelBasis}
 Let $t_1\geq 1$, $t_2,\dots,t_{s-1}\geq 0$, and $t_s\geq1$ be integers. 
 Let  ${\cH}={\cH}_p^{t_1,t_2,\dots,t_s}$ be the $\Z_p\Z_{p^2}\dots\Z_{p^s}$-additive GH code of type $(\alpha_1,\dots, \alpha_s;t_1,\dots,t_{s})$, with $p$ prime,  such that $\Phi({\cH})$ is nonlinear.  Let $\mathbf{w}_k$ be the $k$th row of $A_p^{t_1,t_2,\dots,t_s}$ and $Q=\lbrace (o(\mathbf{w}_k)/p)\mathbf{w}_k\rbrace_{k=1}^{t_1+\dots+t_s}$. Then,
	$\Phi(Q)$ is a basis of $K(\Phi({\cH}))$ and $\kernel(\Phi({\cH}))=t_1+\dots+t_s$.
\end{corollary}
\bibliographystyle{ieeetr}
\bibliography{Manuscript.bib}

\end{document}